\documentstyle[12pt,epsfig]{article}
\newcommand{\be}{\begin{equation}}
\newcommand{\ee}{\end{equation}}
\def\aprle{\buildrel < \over {_{\sim}}}
\def\aprge{\buildrel > \over {_{\sim}}}
\begin{document}
\topmargin 0pt
\oddsidemargin=-0.4truecm
\evensidemargin=-0.4truecm
\renewcommand{\thefootnote}{\fnsymbol{footnote}}
\newpage
\setcounter{page}{1}
\begin{titlepage}     
\vspace*{-2.0cm}
%%%\vspace*{-1.0cm}
\begin{flushright}
FISIST/15-99/CFIF \\
%%\vspace*{-0.2cm}
hep-ph/9909217
\end{flushright}
\vspace*{0.5cm}
\begin{center}
{\Large \bf Small entries of neutrino mass matrices} \\ 
\vspace{1.0cm}

{\large E. Kh. Akhmedov
\footnote{On leave from National Research Centre Kurchatov Institute, 
Moscow 123182, Russia. E-mail: akhmedov@gtae2.ist.utl.pt}}\\
\vspace{0.05cm}
{\em Centro de F\'\i sica das Interac\c c\~oes Fundamentais (CFIF)} \\
{\em Departamento de Fisica, Instituto Superior T\'ecnico }\\
{\em Av. Rovisco Pais, P-1049-001 Lisboa, Portugal}\\
\end{center}
\vglue 0.8truecm
\begin{abstract}
We consider phenomenologically allowed structures of the neutrino 
mass matrix in the case of three light neutrino species. Constraints 
from the solar, atmospheric and reactor neutrino experiments as well 
as those from the neutrinoless double beta decay are taken into 
account. Both hierarchical and quasi-degenerate neutrino mass cases 
are studied. Assuming maximal $\nu_\mu-\nu_\tau$ mixing we derive 
simple approximate expressions giving the values of the neutrino 
masses and remaining lepton mixing angles in terms of the entries 
of the neutrino mass matrix. Special attention is paid to the small 
entries which are usually not specified in discussions of the neutrino 
mass matrix textures. We specifically discuss the stability of 
neutrino masses and lepton mixing angles with respect to the choice 
and variations of these small entries.

\end{abstract}
%\vspace{1.cm}
%%\centerline{Pacs numbers: 14.60.+Pq, 26.65.+t} 
%\vspace{.5cm}
%%\centerline{Keywords: neutrinos, earth, matter effects} 
%\vspace{.5cm}
%\vspace{.3cm}
\end{titlepage}
\renewcommand{\thefootnote}{\arabic{footnote}}
\setcounter{footnote}{0}
\newpage
\section{Introduction}

Recent data from the Super-Kamiokande experiment \cite{SK-AN,SK-SN,Ringberg} 
give a strong support to the interpretation of the atmospheric \cite{ANA} 
and solar \cite{SNP} neutrino problems in terms of neutrino oscillations. 
The results of the solar and atmospheric neutrino experiments along with the 
constraints from the reactor and accelerator experiments 
provided us with an important information on neutrino masses and mixing. 
This has resulted in an increased theoretical activity in the field with a 
large number of publications dedicated to the reconstruction of the neutrino 
masses and mixing and building theoretical models capable of producing the 
requisite neutrino mass matrices 
(see e.g. \cite{recent,Barb,bimax,rad,reviews}). 

In most of these studies some special structures or textures are suggested 
for the neutrino mass matrix. The special structures contain a number of 
large entries and also some small entries which are replaced by zeros in the 
textures. It is usually assumed that this form of the neutrino mass matrices 
emerge as a result of a flavour symmetry. 
In the exact symmetry limit some entries of the neutrino mass matrix vanish, 
but they are supposed to be filled in with small nonzero terms after a small 
breaking of the flavour symmetry is allowed. In many studies, however, these 
small entries of the neutrino mass matrices are not specified or are given 
only by an order of magnitude. 

While the neutrino mass textures reproduce correctly the gross features of the 
neutrino mass matrix, the small terms are of crucial importance for obtaining 
a realistic pattern of neutrino masses and mixing. This includes not only 
reproducing the correct values of the neutrino mass squared differences and 
mixing angles but also the correct correspondence  between the 
ordering of the neutrino masses and the values of the mixing angles.  
Therefore any realistic neutrino mass model has to address the issue 
of the small entries of the neutrino mass matrices.

In the present paper we consider the phenomenologically allowed forms  
of the neutrino mass matrices in the case of three light neutrino species, 
paying special attention to the small entries. 
The analysis is performed without resorting to any specific model of neutrino 
mass and therefore is quite general. Both hierarchical and quasi-degenerate 
neutrino mass cases are considered. Assuming maximal $\nu_\mu$-$\nu_\tau$ 
mixing, as suggested by the Super-Kamiokande atmospheric neutrino data, 
we derive simple approximate formulas which give the values
of neutrino masses and remaining mixing angles in terms of the entries of the
neutrino mass matrix. They allow one to find the neutrino masses and mixings 
off hand, without performing an explicit diagonalization of the neutrino mass 
matrix in each particular case. We discuss the stability of neutrino masses 
and mixing with respect to the choice and variations of the small entries of 
the neutrino mass matrices. The question of stability is an important issue in 
discussions of the radiative corrections in various neutrino mass models 
\cite{rad}. In the present paper it is studied purely phenomenologically, 
and the results can be used in the analyses of any models with three light 
neutrinos species. 

\section{Experimental input}
Currently there are three kinds of experimental indications of nonzero
neutrino mass. These come from the solar neutrino experiments 
\cite{SNP,SK-SN}, atmospheric neutrino experiments \cite{ANA,SK-AN} and 
the accelerator LSND data \cite{LSND}. The LSND result is the only one that 
has not yet been independently confirmed and we therefore choose not to 
consider it here. The results of the solar and atmospheric neutrino 
experiments can then be described in terms of only three light neutrino 
species, $\nu_e$, $\nu_\mu$ and $\nu_\tau$, without introducing sterile 
neutrinos.   

The solar neutrino deficit can be explained through the $\nu_e\to \nu_x$ 
oscillations where $\nu_x$ is $\nu_\mu$, $\nu_\tau$, or their 
linear combination. There are three main types of solutions of the solar 
neutrino problem: small mixing angle MSW effect (SMA), large mixing angle MSW 
effect (LMA) and vacuum neutrino oscillations (VO). 
The atmospheric neutrino data can be explained through the $\nu_\mu \to 
\nu_\tau$ oscillations with possible $\nu_\mu \leftrightarrow \nu_e$ 
oscillations as a subdominant channel. 
The corresponding allowed values of the neutrino parameters can be summarized 
as follows \cite{BKS,Ringberg}: 
\begin{eqnarray}
{\rm SMA:} \quad\quad &\Delta m_\odot^2\simeq (4 - 10)\cdot 10^{-6}
~{\rm eV}^2\,,\quad\quad &\sin^2 2\theta_\odot\simeq (0.1 - 1.0)\cdot 
10^{-2} \nonumber \\
{\rm LMA:} \quad\quad &\Delta m_\odot^2\simeq (2 - 20)\cdot 10^{-5}
~{\rm eV}^2\,,\quad\quad &\sin^2 2\theta_\odot \simeq 0.65 - 0.97\nonumber\\
{\rm VO:} \quad\quad &~~\Delta m_\odot^2 \simeq (0.5 - 5)\cdot 10^{-10} 
~{\rm eV}^2\,,\quad\quad &\sin^2 2\theta_\odot \simeq 0.6 - 1.0 \nonumber \\
{\rm Atm:} \quad\quad & ~\Delta m_{atm}^2 \simeq (2 - 6)\cdot 10^{-3}
~{\rm eV}^2\,,\quad\quad &\sin^2 2\theta_{atm} \simeq 0.82 - 1.0 
\label{constr1}
\end{eqnarray}
We shall use the parametrization of the unitary $3\times 3$ matrix $U$ 
describing the lepton mixing which coincides with the standard quark
mixing matrix parametrization \cite{PDG}. One can then identify the 
lepton mixing angles in (\ref{constr1}) as $\theta_\odot=\theta_{12}$,
$\theta_{atm}=\theta_{23}$. 

In addition to the results listed above, there is an important limit on the 
element $U_{e3}$ of the lepton mixing matrix coming from the CHOOZ reactor 
neutrino experiment \cite{CHOOZ2}, $|U_{e3}|^2 (1-|U_{e3}|^2)<0.045 - 0.02$ 
for $\Delta m_{31}^2\equiv \Delta m_{atm}^2$ ranging in the Super-Kamiokande 
allowed region $(2 - 6)\cdot 10^{-3}$ eV$^2$, which together with the solar 
neutrino observations means 
\be
\sin^2 \theta_{13}\equiv |U_{e3}|^2 \le (0.047 - 0.02)\, \quad 
{\rm for}\quad\Delta m_{31}^2
=(2 - 6)\cdot 10^{-3}~{\rm eV}^2\,.
\label{chooz}
\ee
In the case of quasi-degenerate neutrinos we shall be also using the 
limit on the effective Majorana mass term $m_{eff}$ of $\nu_e$ which comes 
from the Heidelberg-Moscow double beta decay experiment \cite{HM}: 
\be 
m_{eff}\aprle 0.2~{\rm eV}\,.
\label{HM}
\ee
This effective mass is essentially the $m_{ee}$ entry of the 
neutrino mass matrix. 

The results of the existing neutrino experiments can therefore be summarized 
as follows. There are two distinct mass squared difference scales, 
$\Delta m^2_\odot$ and $\Delta m^2_{atm}$, with the hierarchy 
$\Delta m^2_\odot \ll \Delta m^2_{atm}\sim 10^{-3}$ eV$^2$. The mixing angle 
$\theta_{23}$ responsible for the dominant channel of the atmospheric neutrino 
oscillations is close to the maximal one ($45^\circ$), whereas the mixing angle 
$\theta_{12}$ which governs the solar neutrino oscillations can be either small 
or large; for the VO solution it can also be very close or equal to the maximal 
one.  The mixing angle $\theta_{13}$ responsible for
the long-baseline 
$\nu_e\leftrightarrow \nu_x$ oscillations and the subdominant 
$\nu_\mu\leftrightarrow \nu_e$ oscillations of atmospheric neutrinos is either 
small or zero.

\section{Neutrino mass matrices}

The experimental information on the neutrino masses and lepton mixing
angles allows one to reconstruct the phenomenologically allowed form of the 
neutrino mass matrix. We shall adopt the following convention. The neutrino 
mass eigenstate separated from the other two by the large $\Delta m^2_{atm}$ is 
$\nu_3$ whereas those responsible for the solar neutrino problem which are 
separated by the small $\Delta m_\odot^2$ are $\nu_1$ and $\nu_2$. 
In other words, $\Delta m^2_{atm}\simeq \Delta m^2_{31}\simeq \Delta 
m^2_{32}$, $\Delta m_\odot^2 = \Delta m^2_{21}$. 
Neutrino mass spectrum can be either hierarchical ($m_1, m_2\ll m_3$ or $m_1, 
m_2\gg m_3$) or quasi-degenerate, $m_1\approx m_2\approx m_3$, with only mass 
squared differences being hierarchical. The ordering of the $\nu_1$ and $\nu_2$ 
states is unimportant in the case of VO solution of the solar neutrino problem 
but it does make a difference in the case of the SMA and LMA solutions: the 
lower-mass state must have a larger $\nu_e$ component. This condition will 
prove to be very important in deriving the constraints on the entries of the 
neutrino mass matrices. We shall disregard possible CP violation effects in 
the leptonic sector and assume the neutrino mass matrix to be real. Its 
eigenvalues, which we hereafter denote $m_i$ ($i=1,2,3$), can be of either 
sign, depending of the relative CP parities of neutrinos. The physical 
neutrino masses are $|m_i|$. 

We shall be assuming that the mixing angle responsible 
for the atmospheric neutrino oscillations $\theta_{23}=45^\circ$ which is the 
best fit value of the Super-Kamiokande data \cite{Ringberg}. The case when 
$\theta_{23}$ is close $45^\circ$ but not exactly equal to this value can 
be treated similarly. In the first order in the small $\sin 
\theta_{13}\equiv \epsilon$ the lepton mixing matrix takes the form   
%\be
%U_1 = 
%\left(\begin{array}{ccc}
%c      & s     & \epsilon \\
%-\frac{s+c \epsilon}{\sqrt{2}}  & \frac{c-s \epsilon}{\sqrt{2}}  
%& \frac{1}{\sqrt{2}}    \\   
%\frac{s-c \epsilon}{\sqrt{2}}  
% & -\frac{c+s \epsilon}{\sqrt{2}}  
%& \frac{1}{\sqrt{2}}  
%\end{array}
%\right) ~~ ,~~
%\label{U2}
%\ee 
\be
U = \left(\begin{array}{ccc}
c      & s     & \epsilon \\
-\frac{1}{\sqrt{2}} (s+c \epsilon)
& \frac{1}{\sqrt{2}}  (c-s \epsilon)
& \frac{1}{\sqrt{2}}    \\   
\frac{1}{\sqrt{2}} (s-c \epsilon)
 & -\frac{1}{\sqrt{2}} (c+s \epsilon)
& \frac{1}{\sqrt{2}}  
\end{array}
\right)\,,
\label{U1}
\ee 
where $c\equiv c_{12}$, $s\equiv s_{12}$, and the neutrino flavour basis 
is $(\nu_e,\,\nu_\mu,\,\nu_\tau)$. 
The neutrino mass matrix in 
the basis where the charged lepton mass matrix has been diagonalized 
can be written as $m_L =  U\, {\rm diag} (m_1,~m_2,~m_3)\, U^T$ which 
gives 
\be
m_L= \left(\begin{array}{ccc}
 \mu  &\frac{1}{\sqrt{2}}[\epsilon(m_3-\mu)+m_-c s] & 
\frac{1}{\sqrt{2}}[\epsilon(m_3-\mu)-m_-c s] \\
\frac{1}{\sqrt{2}}[\epsilon(m_3-\mu)+m_-c s] & 
\frac{1}{2}(m_3+\mu'-2m_-cs \epsilon) & \frac{1}{2}(m_3-\mu') \\
\frac{1}{\sqrt{2}}[\epsilon(m_3-\mu)-m_-c s] & \frac{1}{2}(m_3-\mu') &
\frac{1}{2}(m_3+\mu' + 2m_-cs \epsilon) 
\end{array}
\right)\,.
\label{mL}
\ee
%%%%%%%%%%
%$m_L$ is the symmetric matrix with the entries 
%\be
%(m_L)_{11}=m_1 c^2+m_2 s^2 
%%\nonumber 
%%\\
%\label{m11}
%\ee
%\be
%(m_L)_{12}=\frac{1}{\sqrt{2}}[(m_3-m_1 c^2-m_2 s^2)\epsilon +(m_2-m_1)c s ] 
%%\nonumber 
%%\\
%\label{m12}
%\ee
%\be 
%(m_L)_{13}=\frac{1}{\sqrt{2}}[(m_3-m_1 c^2-m_2 s^2)\epsilon -(m_2-m_1)c s ] 
%%\nonumber 
%%\\ 
%\label{m13}
%\ee
%\be
%(m_L)_{22}=\frac{1}{2}[m_3+m_1 s^2+m_2 c^2-2 c s 
%\epsilon (m_2-m_1)] 
%%\nonumber 
%%\\ 
%\label{m22}
%\ee
%\be
%(m_L)_{23}=\frac{1}{2}[m_3-(m_1 s^2+m_2 c^2)] 
%%\nonumber 
%%\\ 
%\label{m23}
%\ee
%\be
%(m_L)_{33}=\frac{1}{2}[m_3+m_1 s^2+m_2 c^2+2 c s \epsilon (m_2-m_1)] 
%\label{m33}
%%\label{entries}
%%\end{flushleft}
%\ee
%%%%%%%%%
Here 
\be
\mu=m_1 c^2+m_2 s^2\,,\quad \mu'=m_1 s^2+m_2 c^2\,,\quad m_{-}=m_2-m_1 \,,
\label{entries}
\ee
%%%%%%%%%
We shall be using eq. (\ref{mL}) to derive the phenomenologically allowed 
forms of the neutrino mass matrix for various neutrino mass hierarches. 

By making use of eqs. (\ref{mL}) and (\ref{entries}), which are valid up to 
the small corrections of the order $\epsilon^2 \aprle 0.03$, one can readily 
derive the neutrino mass eigenvalues and lepton mixing angles in terms of 
the entries $m_{ij}$ of the neutrino mass matrix $m_L$:
%%%
%\be
%m_{1,2}=\frac{1}{2}\left\{\frac{m_{22}+m_{33}}{2}+m_{11}-m_{23} 
%\pm \sqrt{\left[\frac{m_{22}+m_{33}}{2}-m_{11}-m_{23}
%\right]^2+2(m_{12}-m_{13})^2}\right\}\,,
%\label{m1,2}
%\ee
%%%%%%%%%%%%%%
%\be
%m_{1,2}=\frac{m_{22}+m_{33}}{4}+\frac{m_{11}-m_{23}}{2} 
%\pm \sqrt{\left[\frac{m_{22}+m_{33}}{4}-\frac{m_{11}+m_{23}}{2}
%\right]^2+\frac{(m_{12}-m_{13})^2}{2}}\,,
%\label{m1,2}
%\ee
%%%%%%%%%%%%%%%%%%%%%%%%%%
\be
m_{1,2}=\frac{1}{2}\left\{(m_{22}+m_{33})/2+m_{11}-m_{23} 
\pm \sqrt{\left[(m_{22}+m_{33})/2-m_{11}-m_{23}
\right]^2+2(m_{12}-m_{13})^2}\right\},
\label{m1,2}
\ee
%%%%%%%%%%%%%%
\be
m_3=\frac{1}{2}(m_{22}+m_{33}+2m_{23})\,,
\label{m3}
\ee
\be
\tan 2\theta_{12}=
%\frac{2cs}{c^2-s^2}=
\frac{\sqrt{2}\,(m_{12}-m_{13})}
{\frac{1}{2}(m_{22}+m_{33})-(m_{11}+m_{23})}\,,
\quad\quad
\epsilon=\frac{\sqrt{2}\,(m_{12}+m_{13})}{m_{22}+m_{33}+2m_{23}-2m_{11}}\,.
%\label{epsilon}
\label{theta12}
\ee
We shall now consider various cases of interest.

\subsection{Hierarchy $|m_1|, |m_2|\ll |m_3|$}

In this case the neutrino mass matrix (\ref{mL}) takes the form 
\footnote{Up to possible trivial sign changes due to the rephasing of the 
neutrino fields.}
\be
m_L = m_0 \left(\begin{array}{ccc}
\kappa      & \varepsilon     & \varepsilon' \\
\varepsilon & ~1+\delta-\delta' & 1-\delta \\   
\varepsilon' & ~1-\delta & 1+\delta+\delta' 
\end{array}
\right)\,
\label{mL2}
\ee 
with small $\kappa$, $\varepsilon$, $\varepsilon'$, $\delta$ and $\delta'$. 
One can now use eqs. (\ref{m1,2})-(\ref{theta12}) to find neutrino masses 
and mixing angles up to the corrections of the order $\epsilon^2$. It is not 
difficult, however, to find more accurate expressions, which include terms 
${\cal O}(\epsilon^2)$, directly from the matrix (\ref{mL2}) 
\footnote{While (\ref{mL}) is valid only up to and including the terms 
${\cal O}(\epsilon)$, the form of the mass matrix (\ref{mL2}) is generic  
for the case $|m_1|, |m_2|\ll |m_3|$, $|\epsilon| \ll 1$.}:
\begin{eqnarray}
m_{1,2} & \simeq & 
%m_0 \left\{ 
\left(\frac{\kappa}{2}-\frac{\varepsilon^2+
\varepsilon'^{2}}{4}\right)+\left(\delta-\frac{\delta'^{2}}{4}\right)
%\right. 
\nonumber \\
& \pm & 
%\left. 
\sqrt{\left[\left(\frac{\kappa}{2}-\frac{\varepsilon^2+
\varepsilon'^{2}}{4}\right)-\left(\delta-\frac{\delta'^{2}}{4}\right)
\right]^2 +\frac{1}{2}\left[(\varepsilon-\varepsilon')+\frac{\delta'}{2}
(\varepsilon+\varepsilon')\right]^2} \,,
%\, \right\}
\label{m12a}
\end{eqnarray}
\be
m_3\simeq 2 +\frac{\varepsilon^2+\varepsilon'^{2}+\delta'^{2}}{2}\simeq 2\,, 
\label{masses}
\ee
\be
\tan 2\theta_{12} \simeq \frac{(\varepsilon-\varepsilon')+\frac{\delta'}{2} 
(\varepsilon+\varepsilon')}
{\sqrt{2}\left[\left(\delta-\frac{\delta'^{2}}{4}\right)-\left(\frac{\kappa}
{2}-\frac{\varepsilon^2+\varepsilon'^{2}}{4}\right)\right]}\,,\quad\quad\quad
\sin \theta_{13}\equiv \epsilon \simeq
\frac{\varepsilon+\varepsilon'}{2\sqrt{2}(1-\kappa/2)}\,. 
\label{thetaepsilon}
\ee
where the eigenvalues $m_i$ are given in units of $m_0$. The value of $m_0$ 
can be fixed through $\Delta m_{31}^2\simeq (2m_0)^2\simeq \Delta m_{atm}^2$. 
The mixing angle $\theta_{13}$ is small by construction and the limit 
(\ref{chooz}) can be easily satisfied. The parameters $\Delta m_{21}^2= 
\Delta m_\odot^2$ and $\theta_{12}$ are fully determined by the small entries 
of the neutrino mass matrix, the values of which should be chosen in accordance 
with the assumed solution of the solar neutrino problem. 

For the MSW solutions 
%of the solar neutrino problem 
(both SMA and LMA), the
lower-mass state out of the two eigenstates, $\nu_1$ and $\nu_2$, must have 
the larger $\nu_e$ component. The condition for this is 
\be  
| 4\delta-\delta'^{2}|> | 2{\kappa}-(\varepsilon^2+
\varepsilon'^{2})|\,.
\label{cond2}
\ee
Therefore the points in the parameter space at which the inequality in 
eq. (\ref{cond2}) is replaced by the equality are unstable: a small change 
of the values of the small parameters can affect drastically the 
pattern of neutrino masses and mixing. One of these points,  
$4\delta-\delta'^{2}=2\kappa-(\epsilon^2+\epsilon'^{2})$, is the point 
at which the denominator in the expression for $\tan 2\theta_{12}$ in 
eq. (\ref{thetaepsilon}) develops a pole, i.e. the mixing becomes maximal. 
Although at the other point, $4\delta-\delta'^{2}=-[2\kappa-(\epsilon^2 
+\epsilon'^{2})]$, the value of $\tan 2\theta_{12}$ remains finite, it is 
an instability point as well.  
This can be explained as follows. At this point 
one has $m_2=-m_1$ so that the eigenstates $\nu_1$ and $\nu_2$ are degenerate. 
Crossing this point in the parameter space would interchange the relative 
position of the two eigenstates -- the lower mass eigenstate would become the 
higher mass one and vice versa. At the same time, their flavour composition 
remains nearly constant since both $\theta_{13}$ and $\theta_{12}$ are regular 
at this instability point. This means that the condition that the lower-mass 
state out of the two eigenstates have the larger $\nu_e$ component is violated 
on one of the sides of this instability point. 

It should be noted that the expression $4\delta-\delta'^{2}$ on left hand side 
of eq. (\ref{cond2}) is the determinant of the $2\times 2$ matrix in the 
23 subspace of the matrix (\ref{mL2}). Therefore in models in which this 
subdeterminant exactly vanishes the condition (\ref{cond2}) is not satisfied 
and the MSW solutions of the solar neutrino problem are not possible. 

\subsection{Inverted hierarchy, $|m_3|\ll |m_1|\approx |m_2|$}
 
In the case $|m_3|\ll |m_1|,|m_2|$ the neutrino mass matrix (\ref{mL}) takes 
the form 
%\be
%m_L = \tilde{m}_0 \left(\begin{array}{ccc}
%\tilde{\kappa}      & \tilde{\varepsilon}     & \tilde{\varepsilon}' \\
%\tilde{\varepsilon} & 1+\tilde{\delta}-\tilde{\delta}' & -1-\tilde{\delta} \\   
%\tilde{\varepsilon}' & -1-\tilde{\delta} & 1+\tilde{\delta}+\tilde{\delta}' 
%\end{array}
%\right)\,
%\label{mL2}
%\ee 
\be
m_L = m_0 \left(\begin{array}{ccc}
\kappa      & \varepsilon     & \varepsilon' \\
\varepsilon & ~1+\delta-\delta' & -1+\delta \\   
\varepsilon' & -1+\delta & ~1+\delta+\delta' 
\end{array}
\right)\,,
\label{mL3}
\ee 
and the requirement $|m_1|\approx |m_2|$ leads to some additional constraints.
%Using eqs. (\ref{m1,2}) and (\ref{m3}) one can find 
There are essentially three possibilities: 
%%%
%(1) $\varepsilon$, $\varepsilon'$ $\delta$ and $\delta'$ small, $\kappa
%\approx 2$; (2) $\varepsilon$, $\varepsilon'$ $\delta$ and $\delta'$ small, 
%$\kappa \approx -2$ (or $\varepsilon\approx -\varepsilon'$, $|\varepsilon|
%\aprle 1$, $\delta$ and $\delta'$ small, $\kappa \approx -2$), and 
%(3) $\varepsilon'\simeq \pm \varepsilon$, $|\varepsilon|\gg 1$, $|\delta|, 
%|\delta'|,|\kappa| \aprle 1$. Case (3) can also be formulated as 
%$\varepsilon\simeq \pm \varepsilon'={\cal O}(1)$, the rest of the entries 
%of the matrix $m_L$ being small. Case (1) leads to $m_1\approx m_2$ whereas 
%cases (2) and (3) yield $m_2 \approx -m_1$. We shall now consider these 
%cases in turn. 

(1) $\varepsilon$, $\varepsilon'$ $\delta$ and $\delta'$ small, $\kappa
\approx 2$. This gives $m_1\simeq m_2$. The neutrino mass eigenvalues 
and mixing angles are given by
\be
m_{1,2}\simeq m_0\left[1+\frac{\kappa}{2}\pm
\sqrt{\left(1-\frac{\kappa}{2}\right)^2+
\frac{(\varepsilon-\varepsilon')^2}{2}}\right]\,, \quad\quad 
m_3\simeq \left(2\delta-\frac{\delta'^{2}}{2}\right) m_0\,, \quad\quad
\label{m12m3}
\ee
\be
\tan 2\theta_{12}\simeq\frac{(\varepsilon-\varepsilon')}
{\sqrt{2}(1-\kappa/2)}\,, \quad \quad\quad 
\sin\theta_{13}=\epsilon\simeq \frac{\varepsilon+
\varepsilon'}{2\sqrt{2}(\delta-\kappa/2)}\,. \quad \quad
\label{thetaepsilon1}
\ee

(2) $\varepsilon$, $\varepsilon'$ $\delta$ and $\delta'$ small, $\kappa 
\approx -2$. This gives $m_2\simeq -m_1$. Formulas (\ref{m12m3}) and 
(\ref{thetaepsilon1}) apply in this case, too. 

The case $m_2\simeq -m_1$ obtains also when $\varepsilon$ and $\varepsilon'$ 
are not small provided they are nearly negative of each other. In this case 
($\varepsilon'\approx -\varepsilon$, $|\varepsilon| \aprle 1$, $\delta$ and
$\delta'$ small, $\kappa \approx -2$) the values of $m_{1,2}$ and 
$\tan 2\theta_{12}$ are again given by the first equations in (\ref{m12m3}) 
and (\ref{thetaepsilon1}), whereas for $m_3$ and $\sin \theta_{13}$ one 
obtains 
\be
m_3\simeq \frac{(\varepsilon+\varepsilon')^2+(\varepsilon-\varepsilon')^2 
\delta+(\varepsilon^2-\varepsilon'^{2})\delta'-4\kappa\delta}
{(\varepsilon^2+\varepsilon'^{2})-2\kappa(1+\delta)-4\delta}\, m_0\,, 
\label{m3new}
\ee
\be
\sin\theta_{13} \simeq -\frac{(\varepsilon+\varepsilon')-\delta(\varepsilon-
\varepsilon')-\varepsilon'(\delta'+m_3/m_0)}{\sqrt{\varepsilon^2 
(\varepsilon^2+\varepsilon'^{2})-2\varepsilon(\varepsilon-\varepsilon')
(\kappa-m_3/m_0)+2(\kappa-m_3/m_0)^2}}\,.
\label{epsnew}
\ee
In both cases (1) and (2) the value of $m_0$ can be fixed through 
$m_1^2\simeq m_2^2\simeq (2m_0)^2\simeq \Delta m_{atm}^2$, whereas 
$\Delta m_{21}^2= \Delta m_\odot^2\simeq 8m_0^2 r$ where $r$ is the square 
root that appears in the first equation in (\ref{m12m3}). The mixing angle 
$\theta_{13}$ is small by construction; in case (1) the value of $\tan 
2\theta_{12}$ depends on how close $\kappa$ is to 2 as compared to 
$\varepsilon-\varepsilon'$; in case (2) $\tan 2\theta_{12}$ is small when 
$\varepsilon$ and $\varepsilon'$ are small but can be large in the case 
$\varepsilon'\simeq -\varepsilon$, $|\varepsilon|\aprge 1$. 

The condition that the lower-mass state out of the eigenstates $\nu_1$ and 
$\nu_2$ have the larger $\nu_e$ component, which is important for the 
SMA and LMA solutions of the solar neutrino problem, is 
\be  
| \kappa| < 2\,.
\label{cond3}
\ee
It constrains only a large entry of the mass matrix and not the  small 
entries. This is related to the particular choice of the parametrization 
of the mass matrix in eq. (\ref{mL3}) which is convenient because it 
leads to the very simple expressions (\ref{m12m3}) and (\ref{thetaepsilon1}). 
In general, this condition reads
$2|m_{11}|<|m_{22}+m_{33}-2m_{23}|$, 
%\be
%2|m_{11}|<|m_{22}+m_{33}-2m_{23}|\,,
%\label{cond4}
%\ee 
which includes both small and large entries. The parameter $\kappa$ {\em is}
required to be close to $\pm 2$ in the cases under discussion. A small change 
of its value can drastically alter the pattern of the neutrino masses and 
mixing. As an example, consider the case $\kappa=1.998$, $\varepsilon=5\cdot  
10^{-3}$, $\varepsilon'=1\cdot 10^{-3}$, $\delta=1\cdot 10^{-2}$ and 
$\delta=3\cdot 10^{-3}$. This yields $m_1\simeq 1.996$, $m_2\simeq 2.002$, 
$m_3\simeq 0.02$ (in units of $m_0$), $\sin\theta_{13}\simeq -2.14\cdot  
10^{-3}$, $\cos \theta_{12}\simeq0.816$ ($\sin^2 2\theta_{12}\simeq 0.889$). 
This choice of parameters is suitable for the LMA solution of the solar 
neutrino problem. The fact that $\cos^2 \theta_{12} >1/2$ means that the 
lower-mass state $\nu_1$ has the larger $\nu_e$ component than the higher 
mass state $\nu_2$, as it should. If, however, the value of $\kappa$ is 
increased by just $0.2\%$ while the other parameters are kept intact, the 
eigenstates $\nu_1$ and $\nu_2$ interchange their position so that now 
$m_1>m_2$ (alternatively, one can rename the states in such a way that $m_1$ 
is always smaller than $m_2$ but then $\sin\theta_{12}$ and $\cos\theta_{12}$ 
interchange). This means that the lower-lying of the two mass eigenstates no 
longer has the larger $\nu_e$ component and therefore there is no solution to 
the solar neutrino problem at all. Similar situation takes place in case (2) 
when $\kappa\simeq -2$. 

We have seen that, like in the case of the normal hierarchy $|m_{1,2}|\ll|m_3|$, 
in cases (1) and (2) of the inverted hierarchy 
%$|m_3|\ll |m_1|\simeq |m_2|$ 
there are instability regions which are close to the certain points in the 
parameter space. 
In the case of the normal hierarchy these instability regions are just small 
domains of the full allowed parameter space, and the parameters can in
general be quite far from these domains. On the contrary, in the cases of the 
inverted hierarchy under discussion, the parameter $\kappa$ {\em must} be close 
to $\pm 2$, so that the situation is generically unstable. This is related to 
the fact that the states $\nu_1$ and $\nu_2$ are quasi-degenerate in this case 
while in the case of the normal hierarchy their masses may be hierarchical. 

(3) $\varepsilon'\simeq \pm \varepsilon$, $|\varepsilon|\gg 1$, $|\delta|, 
|\delta'|$; $|\kappa| \aprle 1$. $\delta$ and $\delta'$ need not be small. 
This case also leads to $m_2 \simeq -m_1$. By a rescaling of the parameters it 
can also be formulated as $\varepsilon'\simeq \pm \varepsilon={\cal O}(1)$, the 
rest of the entries of the matrix $m_L$ being small. This case was suggested  
in \cite{Barb} 
on the basis of the approximate $L_e-L_\mu-L_\tau$ symmetry. 
It is more convenient to use a different parametrization 
of the mass matrix in this case:  
\be
m_L = m_0 \left(\begin{array}{ccc}
~~ \varepsilon     & ~~1              & ~\pm 1           \\
~~  1              & ~~\delta         & ~~~~\varepsilon' \\   
  \pm 1            & ~~~\varepsilon'  & ~~~~\delta' 
\end{array}
\right)\,,
\label{mL4}
\ee 
where $\varepsilon$, $\varepsilon'$, $\delta$ and $\delta'$ are small. 
The neutrino masses and mixing angles are now 
\be
m_1\simeq m_0\left\{\frac{1}{4}[2(\varepsilon\pm\varepsilon')+\delta+
\delta']-\sqrt{2}\right\}\,,  \quad 
m_2\simeq m_0\left\{\frac{1}{4}[2(\varepsilon\pm\varepsilon')+\delta+
\delta']+\sqrt{2}\right\}\,, 
\ee
\be
\!\!\!\!\!\!\!\!\!\!\!\!\!\!\!\!\!\!\!
m_3\simeq \frac{m_0}{2}(\delta+\delta' \mp 2\varepsilon')\,,
\quad\quad \sin \theta_{13}\simeq \mp \frac{\delta-\delta'}{2\sqrt{2}}\,.
\ee
where the upper and lower signs refer to the corresponding signs of 
the matrix element $(m_L)_{13}$ in (\ref{mL4}). In this case, too, the 
mixing angle $\theta_{13}$ is automatically small. The 
mixing angle $\theta_{12}=45^\circ$ up to corrections of the second 
order in the small entries of the mass matrix, i.e. only the VO solution 
of the solar neutrino problem is possible 
\footnote{This remains true also in a more general case in which 
$(m_L)_{13}=A$, $(m_L)_{13}=B$ and for $A\sim B$ the mixing angle 
$\theta_{23}\simeq -\arctan(B/A)$ is large but not necessarily maximal.}. 
This is an example of the so-called bimaximal mixing \cite{bimax}. 
Unlike cases (1) and (2) discussed above, this case is very stable with 
respect to variations of the entries of the neutrino mass matrix. 
This is because the maximal mixing $\theta_{12}=45^\circ$ makes the 
ordering of $\nu_1$ and $\nu_2$ unimportant. 

\subsection{Quasi-degenerate neutrinos, 
$|m_1|\simeq|m_2|\simeq|m_3|$}
The quasi-degenerate cases can be classified according to the relative 
signs of the neutrino mass eigenvalues. 

(1) $m_1\simeq m_2 \simeq m_3$. We shall define the mass splittings 
$\tilde{\delta}$ and $\Delta$ through $m_2=m_1+\tilde{\delta}$, $m_3=
m_1+\Delta$, $|\tilde{\delta}|\ll|\Delta|\ll|m_1|$. Then the neutrino 
mass matrix (\ref{mL}) can be written as
\be
m_L=m_1\,diag(1,\,1,\,1)+\frac{\Delta}{2} M\,,
\label{mL5}
\ee
where the matrix $M$ has the same form as the one on the right hand 
side of eq. (\ref{mL2}). Therefore the formulas of sec. 3.1 apply to 
this case up to the obvious modification of the mass eigenvalues.
Zeroth order texture in this case is proportional to the unit matrix. 

(2) $m_1\simeq m_2 \simeq -m_3$. We now denote $m_2=m_1+\tilde{\delta}$, 
$m_3=-m_1+\Delta$. The mass matrix (\ref{mL}) takes the form
%\begin{eqnarray}
\[
m_L = m_1
\left(\begin{array}{ccc}
           1       & -\sqrt{2}\epsilon & -\sqrt{2}\epsilon  \\
 -\sqrt{2}\epsilon &   ~~0             &  -1   \\
 -\sqrt{2}\epsilon &   -1              &  ~~0  
\end{array}\right)  +  \frac{\Delta}{2}
\left(\begin{array}{ccc}
  0                & \sqrt{2}\epsilon  & \sqrt{2}\epsilon  \\
\sqrt{2}\epsilon   &  1                &  1                \\
\sqrt{2}\epsilon   &  1                &  1   
\end{array}\right) 
~~~~~~~~~~~~~~~~~~~~~~~~~~~~~~~~~~~~~~~
%\nonumber  \\
\]
\be
~~~~~~~~~~~~~~~~~~~~~~~~~~~~~~ +~  \frac{\tilde{\delta}}{2}
\left(\begin{array}{ccc}
 2s^2  & \sqrt{2}(cs-\epsilon s^2) & -\sqrt{2}(cs+\epsilon s^2)\\
\sqrt{2}(cs-\epsilon s^2) & c^2-2cs\epsilon &  -c^2   \\
-\sqrt{2}(cs+\epsilon s^2) &  -c^2  & c^2+2cs\epsilon
\end{array}
\right)\,,
\label{mL6}
%\end{eqnarray}
\ee
where the notation is the same as in eq. (\ref{mL}). The zeroth order 
texture is given by the first term in this equation in which $\epsilon$ is 
set equal to zero. Neutrino masses and mixing angles can be expressed 
through the entries of the mass matrix (\ref{mL6}) with the use of 
eqs. (\ref{m1,2})-(\ref{theta12}).

(3) $m_1\simeq -m_2 \simeq -m_3$ 
\footnote{The case $m_1\simeq -m_2 \simeq m_3$ 
can be obtained from the present one by the change of notation 
$\nu_1 \leftrightarrow \nu_2$.}.  
We now denote $m_2=-m_1+
\tilde{\delta}$, $m_3=-m_1+\Delta$. The mass matrix (\ref{mL}) takes 
the form 
\be
m_L=M_1+M_2+M_3\,,
\label{mL7}
\ee
where $M_2$ and $M_3$ coincide with the second and the third terms on the 
right hand side of eq. (\ref{mL6}) respectively, and 
\be
M_1=-m_1 \left(\begin{array}{ccc}
 -(c^2-s^2)  & \sqrt{2}(cs+\epsilon c^2) & -\sqrt{2}(cs-\epsilon c^2)\\
\sqrt{2}(cs+\epsilon c^2) & c^2-2cs\epsilon &  s^2   \\
-\sqrt{2}(cs-\epsilon c^2) &  s^2  & c^2+2cs\epsilon
\end{array}
\right)\,. 
\label{mL8}
\ee
Notice that, unlike in the two previous cases, the zeroth order 
texture (which obtains from (\ref{mL8}) in the limit $\epsilon=0$) 
depends on the mixing angle $\theta_{12}$. This is because the exact 
degeneracy limit now corresponds to $m_2=-m_1$ and not to $m_2=m_1$, 
so that the mixing between $\nu_1$ and $\nu_2$ remains meaningful. 
Therefore the zeroth order texture in this case depends on the choice 
of the solution of the solar neutrino problem. For the SMA solution one 
has to choose $c=1$, $s=0$ or $c=0$, $s=1$; the corresponding zeroth 
order textures are $diag(-1,\, 1,\, 1)$ and the matrix with  
$m_{11}=m_{23}=m_{32}$, the rest of the elements being zero. Small 
nonvanishing $\theta_{12}$ results when the zeroth order textures 
are perturbed. Two further examples are 
\be
\left(\begin{array}{ccc}
  0        &  ~~\sqrt{2}  & -\sqrt{2}  \\
\sqrt{2}   &  ~~1                &  1  \\
\sqrt{2}   &  ~~1                &  1   
\end{array}\right)\,,\quad\quad 
\frac{1}{3}\left(\begin{array}{ccc}
  -1   & ~~2  & -2  \\
 ~~2   & ~~2  & ~~1  \\
  -2   & ~~1  & ~~2   
\end{array}\right)\,, 
\label{examples}
\ee
the first of which corresponds to the bimaximal mixing
($\theta_{12}=\theta_{23}=45^\circ$) and the second is an example 
of the texture with a large but not maximal $\theta_{12}$: 
$s=1/\sqrt{3}$ ($\sin^2 2\theta_{12}=8/9$). 

We shall now consider one of the examples, namely, $c\simeq 0$, 
$s\simeq 1$, in more detail. The neutrino mass matrix for this case 
can be written as 
\be
m_L = 
m_0\left(\begin{array}{ccc}
  1   & \varepsilon      &  \varepsilon'  \\
\varepsilon  &  \delta   & 1+\delta'    \\
\varepsilon' & ~1+\delta' & \rho    
\end{array}
\right)\,
\label{mL9}
\ee
with small $\varepsilon$, $\varepsilon'$, $\delta$, $\delta'$ and 
$\rho$. From eqs. (\ref{m1,2})-(\ref{theta12}) one finds 
\be
m_{1,2}\simeq \frac{m_0}{2}\left\{\left(\frac{\rho+\delta}{2}-\delta'
\right)\mp \sqrt{\left(2+\delta'-\frac{\rho+\delta}{2}\right)^2+
2(\varepsilon-\varepsilon')^2}\right\}\,,
m_3\simeq m_0\left(1+\delta'+\frac{\rho+\delta}{2}\right)
\label{mnew}
\ee  
\be
\tan 2\theta_{12}=-\frac{\varepsilon-\varepsilon'}{\sqrt{2}[1+
\delta'/2-(\rho+\delta)/4]}\,,\quad\quad 
\sin\theta_{13}=\frac{\varepsilon+\varepsilon'}{\sqrt{2}
[(\rho+\delta)/2+\delta']}\,.
\label{angles}
\ee
The mixing angle $\theta_{13}$ depends on a ratio of the small 
parameters. It is not automatically small and care should be taken in order 
to satisfy the constraint (\ref{chooz}). The mixing angle $\theta_{12}$ is 
generically small because the zeroth order texture corresponds to zero 
mixing. Thus, in this case only the SMA solution of the solar neutrino 
problem is possible. The neutrino mass splittings are 
\be
\tilde{\delta}\simeq m_0 \, [(\rho+\delta)/2-\delta']\,,\quad\quad\quad
\Delta\simeq m_0 \, [\rho+\delta]\,.
\label{diff}
\ee
The condition that the lower-mass state out of the eigenstates $\nu_1$ and 
$\nu_2$ have the larger $\nu_e$ component is 
\be  
\delta'>(\rho+\delta)/2\,.
\label{cond4}
\ee
The point at which the inequality is replaced by the 
equality is unstable with respect to small variations of the parameters. 
As follows from (\ref{diff}), the requirement of the correct hierarchy of 
the mass differences, $|\Delta| \gg |\tilde{\delta}|$, is equivalent to 
$\rho+\delta \simeq 2\delta'$.  This means that in the present case the 
parameters are necessarily close to the instability point.  

The value of the mixing angle $\theta_{13}$ as well as those of the neutrino 
mass splittings depend sensitively on the value of $\delta'$. 
In order for $\tilde{\delta}/m_0$ and $\Delta/m_0$ to be small so must be
$\delta'$, which we assume. Notice, however, that $\delta'$ is the difference 
of the two large matrix elements $m_{23}$ and $m_{11}$. This means that a very 
small relative change of these elements can result in a drastic change 
of the whole pattern of neutrino masses and mixing. This situation is typical 
for all cases of quasi-degenerate neutrinos. 

Finally, it should be noticed that if neutrinos are quasi-degenerate, care 
should be taken in order not to violate the limit (\ref{HM}) coming from the
experiment on the neutrinoless double beta decay. It essentially constrains 
the $m_{11}$ entry of the mass matrices. From the discussion in this 
subsection it follows that the common mass scale $m_0$ must satisfy this 
constraint in all quasi-degenerate cases except in the one of bimaximal 
mixing, for which the zeroth order texture is given by the first matrix in 
eq. (\ref{examples}). This case leads to the VO solution of the solar 
neutrino problem.  As follows from the second example in (\ref{examples}), 
the bound on $m_0$ can also be somewhat relaxed in the case of the LMA 
solution of the solar neutrino problem. In all quasi-degenerate cases in 
which the common mass scale has to satisfy the limit (\ref{HM}) 
neutrinos cannot constitute any appreciable part of the dark matter of the 
universe.

\section{Conclusion}
We have studied systematically the phenomenologically allowed structures 
of the neutrino mass matrix in the case of three light neutrino species.
For both hierarchical and quasi-degenerate 
%neutrino mass 
cases we considered 
neutrino mass matrices which can be obtained by perturbing the well known
zeroth order textures. 

Assuming maximal $\nu_\mu$-$\nu_\tau$ mixing we derived simple analytic
formulas which give neutrino masses and the remaining mixing angles in terms 
of the entries of the mass matrices. 
We have checked our approximate analytic expressions by performing the 
numerical diagonalization of neutrino mass matrix and found a very good 
agreement in each case. 

We analyzed the stability of the neutrino masses and mixing angles with
respect to small variations of the entries of the mass matrices, paying 
special attention to the small entries. 
In particular, we have studied the stability of the condition of having a 
larger $\nu_e$ component in the lower-mass state out of the eigenstates 
$\nu_1$ and $\nu_2$ responsible for the solar neutrino oscillations. 
This condition is important for the SMA and LMA solutions of the solar 
neutrino problem. 

We have found that while the normal hierarchy case $|m_1|, |m_2|\ll |m_3|$ 
is in general very stable, the inverted hierarchy case $|m_3|\ll |m_1|\approx 
|m_2|$ is much less stable (except in the case of bimaximal mixing discussed 
in case (3) of sec. 3.2). The quasi-degenerate case suffers from a number of 
instabilities and therefore is the least natural one. 

The author is grateful to G. C. Branco and M. N. Rebelo for useful 
discussions. 
This work was supported by Funda\c{c}\~ao para a Ci\^encia e a Tecnologia 
through the grant PRAXIS XXI/BCC/16414/98 and also in part by the TMR network 
grant ERBFMRX-CT960090 of the European Union.  

%%%%%%%%%%%%%%%%%%%%%%%%%%%%%%%%%%%%%%%%%%%%%%%%

\end{document}